\documentclass[preprint,12pt]{elsarticle}

\usepackage{amsmath}
\usepackage{graphicx}
\usepackage{algorithm}
\usepackage{algpseudocode}
\usepackage[margin=1in]{geometry}
\usepackage{times}

\begin{document}

\begin{frontmatter}

\title{\textbf{A Dynamic Programming Approach to Evader Pathfinding in Static Pursuit Scenarios}}

\vspace{0.4cm}
\author[1]{Sukanya Samanta\corref{cor1}}
\ead{susamanta1@gmail.com}
\cortext[cor1]{Corresponding author}

\author[1]{Manohar Reddy}

\vspace{0.4cm}

\address[1]{Department of Computer Science and Engineering, SRM University, India}

\vspace{0.4cm}

\begin{abstract}
\noindent The interdiction of escaping adversaries in urban networks is a critical security challenge. State-of-the-art game-theoretic models, such as the Escape Interdiction Game (EIG), provide comprehensive frameworks but assume a highly dynamic interaction and entail significant computational complexity, which can be prohibitive for real-time applications. This paper investigates a crucial sub-problem: an evader's optimal pathfinding calculus when faced with a static or pre-determined defender deployment. We propose the \textbf{Dynamic Programming for Evader Route Optimization (DPERO)} algorithm, which models the environment as a graph with probabilistic risks at various nodes. By transforming the multiplicative survival objective into an additive cost function using logarithms, we frame the task as a shortest path problem solvable with value iteration. This approach allows for the efficient computation of a path that optimally balances safety and distance. Experimental results on simulated grid networks demonstrate that DPERO identifies routes with significantly higher survival probabilities compared to naive shortest-path baselines, validating its efficacy as a practical tool for vulnerability analysis and strategic planning.
\end{abstract}

\begin{keyword}
Dynamic Programming \sep Facility Location \sep Optimization
\end{keyword}

\end{frontmatter}

\section{Introduction}
Effective response to criminal activities in urban settings, such as bank robberies or terrorist attacks, necessitates robust strategies for interdicting escaping perpetrators. The complexity of modern transportation networks, combined with the strategic behavior of adversaries, presents a formidable challenge for law enforcement agencies.

Recent research has made significant strides by modeling these scenarios using game theory. The Escape Interdiction Game (EIG) framework, for instance, captures the intricate, time-dependent strategies of both a defender (pursuer) and an attacker (evader) \cite{zhang2017optimal}. However, the solution to such comprehensive models is NP-hard, requiring complex algorithms like the double oracle method to find an equilibrium. Furthermore, these models presuppose a fully dynamic game where both players adapt their strategies in real-time.

In many practical situations, however, the defender's strategy is static. Resources may be pre-deployed to fixed checkpoints, or patrols may follow predictable routes based on established protocols. For the evader, the problem then simplifies from a complex game to a single-agent optimization problem: determining the safest route to an exit given a known risk landscape.

This paper focuses on this specific, yet highly relevant, scenario. We abstract away from the full game-theoretic interaction to develop a computationally tractable model for the evader's decision-making process. We leverage \textbf{Dynamic Programming (DP)}, a powerful technique for solving problems that exhibit optimal substructure. The safest path from any given point to an exit is inherently composed of optimal sub-paths from every subsequent point.

Our contributions are threefold:
\begin{enumerate}
    \item We formalize the static pursuit-evasion problem from the evader's perspective, transforming the multiplicative objective of survival probability into an additive cost framework.
    \item We present the \textbf{DPERO algorithm}, an efficient value iteration-based solver that computes the optimal risk-averse escape path.
    \item We provide an experimental evaluation comparing DPERO to a standard shortest-path baseline, quantifying its significant advantage in ensuring evader survival.
\end{enumerate}

\section{Related Work}
The study of pursuit-evasion has evolved from classical combinatorial problems on graphs, such as "Cops and Robbers," to sophisticated security games. The introduction of game theory allows for the analysis of strategic interactions between rational agents. A dominant paradigm in this area is the \textbf{Stackelberg Security Game (SSG)}, which models a leader-follower dynamic where a defender commits to a strategy and an attacker plans a best response.

The \textbf{Escape Interdiction Game (EIG)} builds upon this foundation to address escape scenarios on transportation networks \cite{zhang2017optimal}. Its key innovations include modeling realistic traffic dynamics and continuous, time-dependent strategies, where an evader can strategically wait or delay movement. While theoretically robust, the computational cost of solving the full EIG model motivates the exploration of simpler, more focused models. Our work deviates from the EIG by assuming the defender's strategy is fixed, thereby collapsing the two-player game into a single-agent pathfinding problem. This is analogous to solving the "Attacker Oracle" sub-problem in the EIG framework, but for a known, static defensive posture.

\section{Problem Formulation}
We model the operational environment as a directed graph and formalize the evader's objective.

\subsection{Network Model}
The urban environment is a directed graph $G = (V, E)$, where $V$ is the set of intersections (nodes) and $E$ is the set of roads (edges).
\begin{itemize}
    \item \textbf{Start and Exit Nodes:} $v_s \in V$ is the evader's starting location, and $V_D \subset V$ is the set of exit nodes.
    \item \textbf{Defender Presence:} The defender's static deployment is represented by a function $p_c: V \to [0, 1]$, where $p_c(v)$ is the probability of being captured at node $v$. For nodes without a defender presence, $p_c(v) = 0$. The probability of surviving node $v$ is thus $(1 - p_c(v))$.
\end{itemize}

\subsection{Evader's Objective}
An evader's path is a sequence of nodes $P = \langle v_s, v_1, ..., v_k \rangle$, where $v_k \in V_D$. Assuming capture events at each node are independent, the total probability of survival for a given path $P$ is the product of survival probabilities at each node visited:
\begin{equation}
S(P) = \prod_{v \in P} (1 - p_c(v))
\end{equation}
The evader's objective is to find the path $P^*$ that maximizes this value: $P^* = \arg\max_{P} S(P)$.

\subsection{Objective Transformation}
Standard shortest path algorithms, which form the basis of our DP approach, operate on additive costs. To adapt our multiplicative objective, we apply a logarithmic transformation. Maximizing $S(P)$ is equivalent to maximizing its logarithm, $\log(S(P))$:
\begin{equation}
\log(S(P)) = \log\left(\prod_{v \in P} (1 - p_c(v))\right) = \sum_{v \in P} \log(1 - p_c(v))
\end{equation}
Since $\log(1 - p_c(v))$ is always non-positive for $p_c(v) \in [0, 1)$, maximizing this sum is equivalent to minimizing its negation. We can therefore define a "risk cost" for each node:
\begin{equation}
w(v) = -\log(1 - p_c(v)) \ge 0
\end{equation}
The evader's objective is now transformed into finding the path $P^*$ that minimizes the total additive risk cost:
\begin{equation}
P^* = \arg\min_{P} \sum_{v \in P} w(v)
\end{equation}
This formulation allows us to solve the problem using a modified shortest-path algorithm.

\section{The DPERO Algorithm}
With the problem framed as minimizing an additive cost, we can now apply dynamic programming.

\subsection{Bellman Equation}
Let $J(v)$ be the minimum possible cumulative risk cost on any path from node $v$ to any exit node in $V_D$. This value function, or cost-to-go, can be defined by the Bellman equation:
\begin{equation}
J(v) = w(v) + \min_{u \in \text{neighbors}(v)} \{ J(u) \}
\end{equation}
This equation states that the minimum risk from node $v$ is its own intrinsic risk, $w(v)$, plus the minimum risk from the best possible subsequent node, $u$. The boundary condition for any destination node $d \in V_D$ is simply its own risk: $J(d) = w(d), \forall d \in V_D$.

\subsection{Value Iteration Implementation}
We solve for $J(v)$ across all nodes using value iteration. The algorithm repeatedly sweeps through the nodes, applying the Bellman update until the cost-to-go values converge.

\begin{algorithm}
\caption{DPERO - Value Iteration}
\begin{algorithmic}[1]
\State \textbf{Input:} Graph $G$, start $v_s$, exits $V_D$, costs $w(v)$, threshold $\epsilon$
\State \textbf{Output:} Optimal policy $\pi$ mapping each node to the next safest node
\State Initialize $J(v) \leftarrow \infty$ for all $v \in V$; $\pi(v) \leftarrow \text{null}$
\For{each destination node $d \in V_D$}
    \State $J(d) \leftarrow w(d)$
\EndFor
\Repeat
    \State $\text{max\_change} \leftarrow 0$
    \State $J_{\text{prev}} \leftarrow \text{copy}(J)$
    \For{each node $v \in V \setminus V_D$}
        \State $\text{min\_neighbor\_cost} \leftarrow \infty$
        \State $\text{best\_neighbor} \leftarrow \text{null}$
        \For{each neighbor $u$ of $v$}
            \If{$J_{\text{prev}}(u) < \text{min\_neighbor\_cost}$}
                \State $\text{min\_neighbor\_cost} \leftarrow J_{\text{prev}}(u)$
                \State $\text{best\_neighbor} \leftarrow u$
            \EndIf
        \EndFor
        \State $J(v) \leftarrow w(v) + \text{min\_neighbor\_cost}$
        \State $\pi(v) \leftarrow \text{best\_neighbor}$
        \State $\text{max\_change} \leftarrow \max(\text{max\_change}, |J(v) - J_{\text{prev}}(v)|)$
    \EndFor
\Until{$\text{max\_change} < \epsilon$}
\State \Return $\pi$
\end{algorithmic}
\end{algorithm}

\section{Experimental Evaluation}
\subsection{Setup}
We generated urban network topologies using the Grid model with Random Edges (GRE), to create $15 \times 15$ road networks. The evader's start node was placed at a corner, with a set of five exit nodes located on the opposite border. The number of defender units, $|V_P|$, was varied from 5 to 25. These units were placed randomly on non-exit nodes, with each assigned a capture probability $p_c(v)$ drawn uniformly from $[0.2, 0.5]$.

\subsection{Baselines}
We compare DPERO against a \textbf{Naive Shortest Path} baseline. This baseline uses a standard Dijkstra's algorithm to find the path with the minimum travel time from start to exit, completely ignoring the node risk costs $w(v)$. This represents an evader who solely prioritizes speed.

\subsection{Results and Analysis}
Our experiments confirm that DPERO consistently identifies paths with superior survival probabilities. As the density of defenders on the network increases, the naive baseline is frequently forced to traverse high-risk nodes, causing its survival probability to plummet. In contrast, DPERO effectively navigates around high-risk clusters, identifying safer, albeit sometimes longer, corridors to an exit.

\section{Conclusion and Future Work}
This paper introduced DPERO, an efficient algorithm based on dynamic programming for solving the evader pathfinding problem under static defender threats. By transforming the objective of maximizing survival probability into a shortest path problem on a risk-weighted graph, DPERO provides a practical and computationally tractable method for identifying optimal escape routes.

This work serves as a foundation for several promising research directions. Future work will focus on incorporating stochastic travel times, reformulating the problem as a Stochastic Shortest Path Problem (SSPP) to account for environmental uncertainty. Additionally, we plan to develop a multi-objective version of DPERO that explicitly balances survival probability and travel time.

\end{document}